\documentclass[a4paper,11pt]{article}
\usepackage{jcappub} 
\usepackage{lineno}

\title{\boldmath Extended Gravity and Black Hole Shadows: Rotation Accounting}

\author[a,b]{Stanislav Alexeyev,}
\author[b]{Oleg Zenin,}
\author[b]{Artem Baiderin,}

\affiliation[a]{Sternberg Astronomical Institute, Lomonosov Moscow State University,\\ Univeresitesky Prospekt, 13, Moscow, 117234, Russia}
\affiliation[b]{Department of Quantum Theory and High Energy Physics, Physics Faculty, Lomonosov Moscow State University,\\ Vorobievi Gory, 1/2, Moscow, 117234, Russia}

\emailAdd{alexeyev@physics.msu.ru}

\abstract{
We obtain black hole rotating solutions for Horndesky theory (specific partial case), bumblebee model and Gauss-Bonnet scalar gravity using the specially improved Newman-Janis algorithm. The shadow profiles for these metrics were calculated. Applying the limitations from the Event Horizon Telescope we find the opportunity to constrain model parameters from considered extended gravity theories. We show that for three considered models two of them (Horndesky theory and Gauss-Bonnet scalar gravity) weaken the effect of rotation and bumblebee model enhances it. This conclusion matches the previously obtained one that extended gravity theories by themselves correct the effect of rotation in both directions.
}

\begin{document}
\maketitle
\flushbottom

\section{Introduction}\label{sec1}

The new data with the increasing accuracy from the Event Horizon Telescope (EHT) \cite{EventHorizonTelescope:2023fox} causes the necessary to improve the theoretical predictions especially on black hole (BH) shadows profiles. First models of BH shadows based on spherically-symmetric space-times \cite{Zakharov:2005bc}. To extend the consideration an additional parameter named tidal charge from Reissner-Nordstrom metric was added \cite{Dadhich:2000am}. It provides an opportunity to measure the new physics contribution  \cite{Zakharov:2014lqa}. Further it was shown that tidal charge and next order perturbative corrections allow to improve the theoretical model for better compliance EHT data in the future \cite{Alexeyev:2018ofs,Alexeyev:2022mqb}. 

From the beginning it was obvious that both studied BHs are rotating ones \cite{EventHorizonTelescope:2022apq,EventHorizonTelescope:2019pgp}. Recently the rotating speed was measured \cite{Cui:2023uyb}. Therefore in order to make the next step in improving of the theoretical predictions one has to account the BH rotation using Kerr-like metrics \cite{Prokopov:2020ewm}. Here it is necessary to note that (in full analogy with GR) the first obtained local metric in each new gravity model is spherically-symmetric because it represents the most simple case. In the same time the direct search of axially-symmetric solution by solving directly Einstein-Hilbert equations appears to be not very easy \cite{Chandrasekhar:1985kt}. That is why the alternative method allowing to generate rotating solutions from non-rotating ones was suggested \cite{Newman:1965tw}. The interest to Newman-Janis algorithm increased during last years and allowed to turn the Newman-Janis algorithm into the maximally algorithmic form \cite{Azreg-Ainou:2014aqa}. In this way recently new relations between a tidal charge and an angular momentum were obtained for GR BHs \cite{Karas:2023uac} and for bumblebee model \cite{Capozziello:2023tbo} and for non-local gravity models \cite{c02}. These papers point out the interesting phenomenon: when the Kerr-like metric with additional parameters (like tidal charge) is considered these new parameters could change the size and the form of the shadow. Hence if (after the increasing of the experimental accuracy) EHT would discover that the shadow does not exactly matches the Kerr metric it would make possible to estimate the tidal charge contribution. It means that the contribution of new physics would be measured. Therefore we study such behaviour in Horndeski model, scalar Gauss-Bonnet gravity and bumblebee model following our previous results on BH solutions from these theories in the non-rotating case \cite{Prokopov:2021lat}. 

Here it is necessary to comment the status of the Newman-Janis algorithm. In general the symmetry group of the rotating BH solution for an arbitrary gravity model could be more complicated than being realised in Newman-Janis algorithm. Moreover it is individual in each case. To obtain the general form of the solution one has to solve the Einstein equations in each case. Using the Newman-Janis algorithm one generates the rotating solution with the symmetry group being subgroup of real solution symmetry group and therefore generates some degenerated solution (like Kerr-like metric in $D>4$ space-time with 1 momenta). As our aim is to estimate the changes on BH shadow profiles when the spinning Kerr-like metric is used. This means that using the simplest form of spinning solution obtained by Newman-Janis algorithm we proceed new estimation being the next step further.

Here it is necessary to note that for Horndeski model we consider only the simplest solution at some particular case, really the amount of BH solutions in Horndeski theory is infinite.

The structure of our paper is the following: Section 2 devoted to the technical details of rotating solutions generation, Section 3 deals with the BHs shadows modelling for each theory mentioned in the previous paragraph and Section 4 contains the discussion and our conclusions.

\section{Rotating Solutions Generation: Technical Details}

Following \cite{Azreg-Ainou:2014aqa} we consider the spherically-symmetric space-times parametrized with the metric functions $G(r)$, $F(r)$ and $H(r)$. Such form is over-defined in a case to cover maximal amount of classes and has the form:
\begin{equation}\label{metric} 
ds^2 =  - G(r) dt^2 + \frac{1}{F(r)} dr^2 + H(r)d\Omega^2.
\end{equation}
The development of the Newman-Janis algorithm reduces the problem to the direct calculation of the new metric components as:
\begin{eqnarray}\label{spin_metri}
g_{tt} & = & -\frac{FH + a^2\cos^2{\theta}}{(K + a^2\cos^2{\theta})^2} \Psi,\qquad   
g_{t\phi} = -a\sin^2{\theta}\frac{K - FH}{(K + a^2\cos^2{\theta})^2}\Psi,\nonumber \\
g_{\theta \theta} & = & \Psi,\qquad g_{rr}=\frac{\Psi}{FH + a^2},\qquad
g_{\phi \phi} = \Psi\sin^2{\theta}(1 + a^2\sin^2{\theta}\frac{2K - FH + a^2\cos^2{\theta}}{(K + a^2\cos^2{\theta})^2}). \nonumber
\end{eqnarray}
Here $K = H(r)\sqrt{\frac{F(r)}{G(r)}}$ and all the components are defined up to the $\Psi(r,y^2,a)$ where $y \equiv \cos{\theta}$. This additional function $\Psi(r,y^2,a)$ must satisfy the following system:
\begin{eqnarray}\label{dif_eq}
&& \lim_{a\to 0} \Psi(r,y^2,a) = H(r), \qquad
(K + a^2y^2)^2(3\Psi_r\Psi_{y^2} -2\Psi\Psi_{r,y^2}) = 3a^2K_r\Psi^2,\nonumber\\
& \Psi & [K^2_r + K(2 - K_{rr}) -a^2y^2(2 + K_{rr})] + (K + a^2y^2)[(4y^2\Psi_{y^2} - K_r\Psi_r] = 0.
\end{eqnarray}
The first condition in Eq (\ref{dif_eq}) means that when $a \to 0$ the solution is non-rotating. This non-rotating solution could be represented in 2 forms ($\Psi_n$ and $\Psi_c$) coupled via conformal transformation so the initial metric is:
\begin{eqnarray}
    ds_{c}^{2} = \Psi_c/\Psi_nds^2_n.
\end{eqnarray}
One looks for the solution of Eq (\ref{dif_eq}) in the form:
\begin{eqnarray}
\Psi_c &=& H(r)\exp{[a^2f(r,a^2y^2,a)]} \approx H(r) + a^2X(y^2,r) + o(a^2),
\end{eqnarray}
where the Taylor expansion is used and 
\begin{eqnarray}\label{new_dif_eq}
&& KH_rK_r + HK_r^2 + HK(K_{rr} -2) = 0, \qquad X(y^2,r) = \frac{H^2(8K - K_{r}^2)y^2}{K^2(8H - H_rK_r)}, \nonumber \\
&& K_r(8K - K_r^2)K_{rrr} + K_r^2(K_{rr} - 2)^2 - 4 KK_{rr}(K_{rr} + 4) + 48K = 0.
\end{eqnarray}
Here $A_r = \partial A/\partial r$. 

\subsection{Horndeski model}

The spherically-symmetric metric we considered earlier is \cite{Babichev:2017guv,Prokopov:2021lat}:
\begin{eqnarray}\label{horn} 
ds^2 =  &-& (1 - \frac{2M}{r} - \frac{8\alpha_{5}\eta}{5r^3})dt^2 + \frac{1}{1 - \frac{2M}{r} - \frac{8\alpha_{5}\eta}{5r^3}}dr^2 + r^2d\Omega^2 ,
\end{eqnarray}
where $\alpha_5$ and $\eta$ are model parameters. After the applying of the Newman-Janis algorithm in the form from the previous section one obtains ($\rho^2=r^2+a^2\cos^2\theta$, all the other metric components vanish):
\begin{eqnarray}\label{spin_metri_horn1}
g_{tt} & = & - \left( 1 - \frac{2Mr}{\rho^2} - \frac{8 \alpha_5 \eta}{5r}\right) ,\qquad 
g_{t\phi} = - \frac{2a\sin^2\theta}{5r\rho^2} \left( 4\alpha_5 \eta + 9 Mr^2 \right),\nonumber \\
g_{rr} & = & \rho^2 \left( - \frac{8\alpha_5\eta}{5r} + a^2 - 2Mr + r^2  \right)^{-1}, \qquad
g_{\theta \theta} = \rho^2 ,\nonumber \\
g_{\phi\phi} & = & \frac{\sin^2\theta}{\rho^2} \biggl(r^4 + 2 a r^2 \cos^2\theta + a^4 \cos^4\theta + \frac{8a^2\alpha_5\eta\sin^2\theta}{5r} \nonumber \\ & + & 2aMr\sin^2\theta + a^2 r^2 \sin^2\theta + a^4 \cos^2\theta\sin^2\theta \biggr),
\end{eqnarray}

\subsection{Bumblebee model}

The action for unique bumblebee field $B_{\mu}$ is \cite{Casana:2017jkc}:
\begin{eqnarray}\label{b1}
S_B & = & \int d^{4}x\mathcal{L}_B=\int d^{4}x(\mathcal{L}_g + \mathcal{L}_{gB} +
\mathcal{L}_K + \mathcal{L}_V \nonumber \\ & + & \mathcal{L}_M),
\end{eqnarray}
where $\mathcal{L}_g$ is usual GR action (with cosmological constant), $\mathcal{L}_{gB}$ is the connection between gravity and bumblebee field, $\mathcal{L}_K$ are kinetic terms of bumblebee field and other self-action ones, $\mathcal{L}_V$ is the potential originating in spontaneous breaking of Lorenz symmetry, $\mathcal{L}_M$ is the matter and its interactions with bumblebee field. Considering the case without torsion and cosmological constant ($\Lambda=0$) one obtains:
\begin{eqnarray}\label{b2}
\mathcal{L}_B & = & \frac{e}{2\kappa}R+\frac{e}{2\kappa}\xi B^{\mu}B^{\nu}R_{\mu\nu}-\frac{1}{4}eB_{\mu\nu}B^{\mu\nu} - eV(B^{\mu})+\mathcal{L}_M ,
\end{eqnarray}
where $e=\sqrt{-g}$ is a constant of the non-minimal coupling between gravity and bumblebee field. 

We start from the metric in the form \cite{Casana:2017jkc,Prokopov:2021lat}:
\begin{eqnarray}\label{metric_bb} 
ds^2 =  &-& (1 - \frac{2M}{r}) dt^2 + \frac{1 + l}{1 - \frac{2M}{r}}dr^2 + r^2d\Omega^2 ,
\end{eqnarray}
where $l$ is the bumblebee parameter.

After applying of the Newman-Janis algorithm it takes the form (as in the previous subsection all the other metric components vanish, $A$-$K$ used only here):
\begin{eqnarray}\label{spin_metri_bb}
g_{tt} & = & \frac{r^{-1+\sqrt{1+l}} A B }{\sqrt{1+l} C D}, \qquad g_{t\phi} = -\frac{ar^{-l+\sqrt{1+l}} E B \sin^2{\theta}}{(1+l)C D},\qquad g_{rr} = \frac{(1+l)r^{-l+\sqrt{1+l}} B}{C G},\nonumber \\
g_{\theta \theta} & = & r^{1+\sqrt{1+l}}+\frac{a^2(-4+8\sqrt{1+l})r^{-l+\sqrt{1+l}}\cos^2{\theta}}{8-2(1+\sqrt{1+l})}, \\
g_{\phi \phi} & = &\frac{r^{-l+\sqrt{1+l}}\sin^2{\theta}(B+5a^2\cos^2{\theta})}{(1+l)C D} \times \Bigl( D(1+l)-Ka^2\cos^2{\theta}\Bigr), \nonumber 
\end{eqnarray}
where 
\begin{eqnarray*}
A & = & (2Mr^{1+l}-r^{1+\sqrt{1+l}}-a^2\cos^2{\theta}-a^2l\cos^2{\theta}), \\
B & = & -3r^2 + \sqrt{1+l} r^2 - 3a^2\cos^2{\theta} - 4a^2\sqrt{1+l}\cos^2{\theta}, \\
C & = & -3+\sqrt{1+l}, \qquad D = r^2+a^2\sqrt{1+l}\cos^2{\theta}, \\
E & = & -r^2-lr^2-2\sqrt{1+l}Mr^{\sqrt{1+l}}+\sqrt{1+l}r^{1+\sqrt{1+l}}, \\
G & = & a^2+a^2l-2Mr^{1+l}+r^{1-\sqrt{1+l}}, \\
F & = & -2Mr^{\sqrt{1+l}}+r^{1+\sqrt{1+l}}-a^2 l\cos^2{\theta}, \qquad K = \sqrt{1+l}F-r-2lr^2-D.
\end{eqnarray*}

\subsection{Scalar Gauss-Bonnet Gravity}

The discussed model includes all second order curvature corrections in the form \cite{Yunes:2011we,Prokopov:2021lat}:
\begin{eqnarray}\label{sgb1}
S & = & \int d^4x\sqrt{-g}\biggl[ \kappa 
 +\alpha_1 f_1 (\vartheta)R^2 + \alpha_2f_2 (\vartheta) R_{ab}R^{ab} + \alpha_3 f_3 (\vartheta) R_{abcd}R^{abcd} \nonumber \\ & + & \alpha_4f_4 (\vartheta) R_{abcd}^* {R^*}^{abcd} 
- \frac{\beta}{2}\Bigl( \nabla_a \vartheta \nabla^a \vartheta +2V (\vartheta) \Bigr) + \mathcal{L}_{mat}\biggr].
\end{eqnarray}
Here $g$ is the determinant of the metrics $g_{ab}$, ($R$, $R_{ab}$, $R_{abcd}$ and $R_{abcd}^*$) are Ricci scalar, tensor, Riemann tensor and its dual one, $\mathcal{L}_{mat}$ is matter Lagrangian, $\vartheta$ is scalar field, ($\alpha_i, \beta$) are coupling constants and $\kappa=(16\pi G)^{-1}$. The spherically-symmetric space-time with the metric functions from Eq (\ref{metric}) is parametrized as follows:
\begin{eqnarray}
G(r) &=& f_s \left(1 + \frac{\xi}{3r^3f_s}\right) + o\left(\frac{1}{r^3}\right), \\
F(r) &=& \frac{f_s}{(1 - \frac{\xi}{r^3f_s})} + o\left(\frac{1}{r^3}\right),\qquad H(r) = 2\frac{K}{K_r}r, \nonumber
\end{eqnarray}
where $f_s = 1 - \frac{2M}{r}$. After applying of the Newman-Janis algorithm the metric takes the form (as previously all the other metric components vanish):
\begin{eqnarray}\label{spin_metri_gb}
g_{tt} & = & \frac{r^2(E+F\cos^2{\theta})}{AB},\qquad g_{t\phi} = -\frac{aCD\sin^2{\theta}}{AB},\nonumber \\
g_{rr} & = &-\frac{AB}{r^2(E+F)},\qquad g_{\theta \theta} = \frac{B}{3r^2}, \qquad g_{\phi \phi} = \frac{T}{3r^2AB},
\end{eqnarray}
where
\begin{eqnarray*}
A & = & \xi+2Mr^2-r^3, \qquad B = 2\xi M+\xi r+3r^4+3a^2r^2\cos^2{\theta}, \qquad C = 2\xi M+\xi r+3r^4, \\
D & = & A+16M^2r^2-16Mr^4+4r^5, \\
E & = & 32\xi M^3r-16\xi M^2r^2-8\xi Mr^3+4\xi r^4 + 48M^2r^5-48Mr^6+12r^7, \\
F & = & -3a^2\xi-6a^2Mr^2+3a^2r^3, \qquad G = 16\xi M^3r^5+2\xi r^6+24M^2r^7-24Mr^8+6r^9, \\
K & = & 2\xi^2M+\xi^2r+4\xi M^2r^2+2\xi r^4+6Mr^6-3r^7, \\
Q & = & 4\xi^3 M(M+r)+\xi^2 r^2(\xi+2M^3+4M^2r + 10Mr^2+5r^3)+3\xi r^6(8M^2+r^2)+9r^{10}(2M-r), \\
T & = & 1 + Q + 9a^4 r^4 A \cos^4{\theta} + 6 a^2 r^2 G \sin^2{\theta} + 9a^4r^4A\cos^2{\theta} \sin^2{\theta}
\end{eqnarray*}

\section{BH Shadow Calculation}

\subsection{Technical Remarks}

The shadow profile is defined by the last stable orbit therefore the aim is to find all the components from the corresponding equations. In Kerr case these values can be found from the isotropic Hamilton-Jacobi equation \cite{Chandrasekhar:1985kt}. So as a first step finds the solution of this equation in the form (where $S$ is Hamilton-Jacobi function):
\begin{equation}\label{eq:HJE}
    g^{\mu \nu} \frac{\partial S}{\partial x^\mu} \frac{\partial S}{\partial x^\nu} =0. 
\end{equation}
Applying conserved values $E=-p_t$ and $L_z=p_\phi$ (energy and angular momentum) we look for a solution in the form:
\begin{equation}
    S = - Et + L_z\phi + S_r(r) + S_\theta(\theta) ,
\end{equation}
After dividing variables:
\begin{eqnarray}
\mathcal{R}(r) & = & \left(\omega + a^2 - a\lambda\right)^2-(f_r^{-1} r^2 +a^2)\left[\eta+\left(a-\lambda\right)^2\right],\nonumber \\
\Theta(\theta) & = & \eta+\cos^2 \theta \left(a^2-\frac{\lambda}{ \sin^2 \theta}\right),
\end{eqnarray}
where $\eta=\frac{Q}{E^2}$, $\lambda=\frac{L_z}{E}$ and $Q$ is Carter constant. To calculate the spherical photon orbit one has to solve the equations: 
\begin{equation}
    \mathcal{R} = 0, \qquad \frac{d\mathcal{R}}{dr}=0.
\end{equation}
As a result one obtains the dependence of $\lambda$ and $\eta$ upon metric functions. Finally it is necessary to consider the plane oriented in the normal direction to far observer. Shadow coordinates on such a plane are: 
\begin{eqnarray}
    x' & = & -\frac{\lambda}{\sin\theta_0}, \label{eq45}\\
    y' & = & \pm \sqrt{\eta+a^2 \cos^2\theta_0 - \frac{\lambda^2}{\tan^2\theta_0}}, \label{eq46}
\end{eqnarray}
where $\theta_0$ is the solid angle between the rotation plane and the observer's sight line, $\lambda$ and $\eta$ defined as:
\begin{eqnarray}
    \lambda & = & \frac{K+a^2}{a}-\frac{2K'}{a}\frac{(FH+a^2)}{(HF)'}, \label{eq47} \\
    \eta & = & \frac{4(a^2+FH)}{\left((HF)'\right)^2}\left(K'\right)^2-\frac{1}{a^2} \Bigr[ K-\frac{2(FH+a^2)}{(HF)'}K' \Bigr] . \label{eq48}
\end{eqnarray}

To calculate the $X$ and $Y$ coordinates on picture's plane using Eqs (\ref{eq45}-\ref{eq46}) we improved our Python code created earlier \cite{Prokopov:2021lat,c02}. In addition such values as $r_s$ (the effective size of BH shadow), $D$ (the shadow shift from the centre), $\delta=\Delta_{cs}/r_s$ (shadow distortion in non-rotating case, $\Delta_{cs}$ is the distance between the left shadow shape and its circular approximation). Earlier the shadow profile for BH from non-local gravity theory was presented \cite{c02} in comparison with Sgr A* \cite{EventHorizonTelescope:2022apq} and M87*  \cite{EventHorizonTelescope:2023fox} data. We consider the most probable configurations of Sgr A*. So the inclination of the rotational plane relative to the observer's sight line is equal to $\pi/6$ and $a=0.5$ and $a=0.94$ (relative mass $M$) \cite{EventHorizonTelescope:2022apq}. For comparison we always show the Schwarzschild-like case $a=0$. Note that the rotation value $a=0.9375$ was obtained from relativistic jet M87* observations  \cite{Cui:2023uyb} so there is the BH with high speed rotation.

\subsection{Horndesky model}

\begin{figure}[htbp]
\centering
\includegraphics[width=\textwidth]{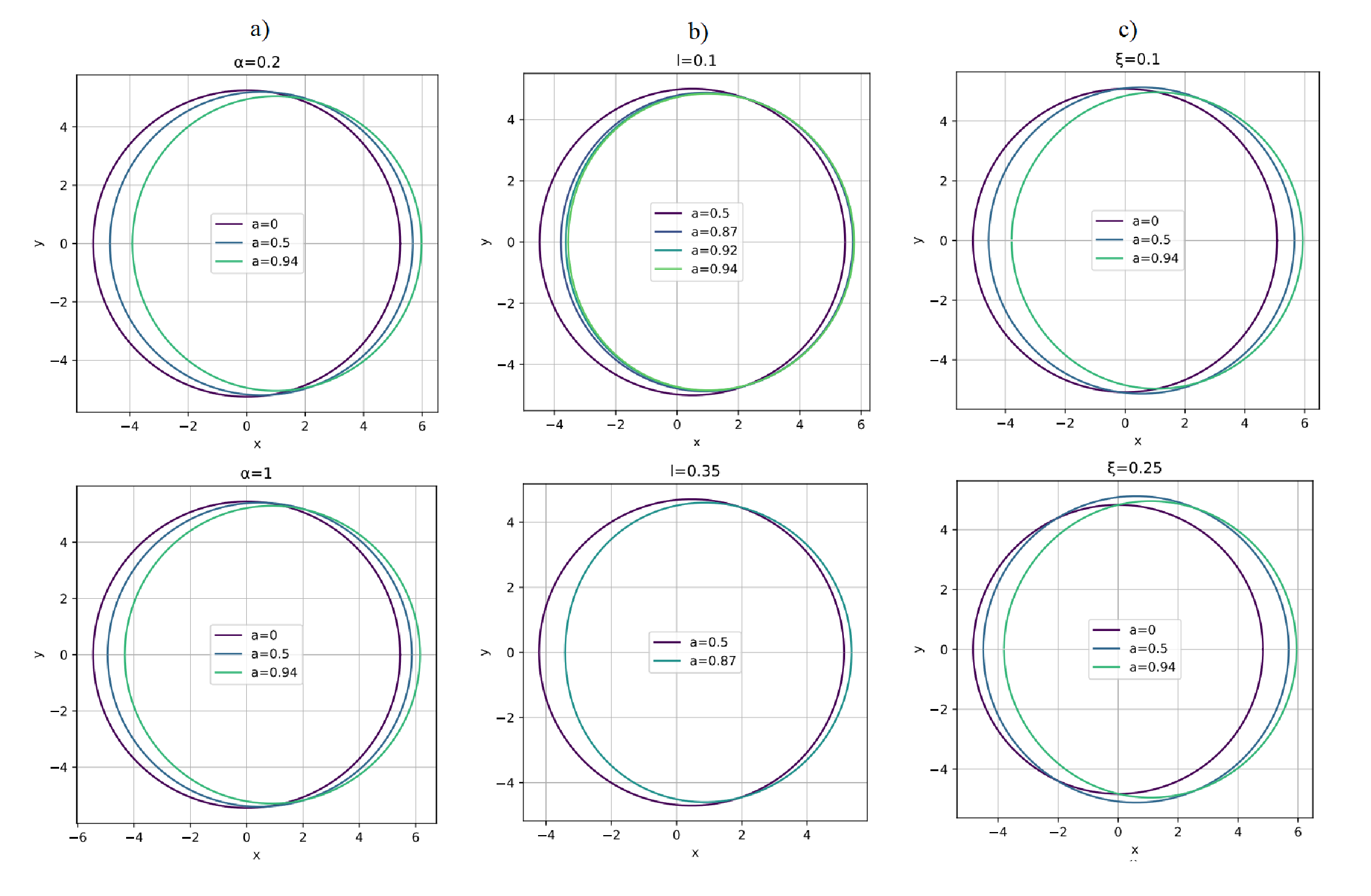}
\caption{BH shadow profiles for different values of rotation parameter $a$ against (a) $\alpha$ in Horndesky model, (b) $l$ in bumblebee model and (c) $\xi$ in scalar Gauss-Bonnet gravity. The minimal and maximal values of additional parameters are established. The inclination angle of the rotational plane is equal to $\theta_0=\frac{\pi}{6}$ (Sgr A*).}
\label{sh}
\end{figure}

In our calculations we apply the obtained Kerr-like metric (\ref{spin_metri_horn1}) introducing new parameter $\alpha = 8\alpha_5\eta/5$. Firstly we obtain the shadow profiles for different values of $\alpha$ (Fig.~\ref{sh}(a)). At the next step we estimate the effective shadow radius (Fig.~\ref{rs}(a)). Coloured region shows the limitations from \cite{EventHorizonTelescope:2022apq} on BH shadow size. Note that increasing of massless angular momentum $a$ causes the decreasing of the shadow size just as in \cite{c02}. The value $\alpha=1$ is excluded when $a=0.5$. Nevertheless at $a=0.94$ the shadow size remains in the allowed region. When $\alpha=0.8$ the configuration with $a=0.94$ also remains applicable. When $\alpha<0.5$ both configurations continue to be allowed. Also note that increasing $\alpha$ causes the increasing of shadow size so $\alpha$ acts opposite to $a$ and in non-rotating case the big values of $\alpha$ are excluded (only $\alpha=1$ remains allowed). Further same as in \cite{c02} the shift $D$ 
grows linearly with $a$ and has no big difference at different $\alpha$ (Fig.~\ref{D_delta}(a up)). 

The last considered parameter is distortion $\delta$ (Fig.~\ref{D_delta}(a down)). As one can see for all $\alpha$ distortion is equal about $0.5\%-1\%$. When $a=0.94$ the distortion increases (once more $\alpha$ acts opposite $a$: as $\alpha$ grows, distortion caused by $a$ becomes less) from $2\%$ at $\alpha=1$ till $5.5\%$ at pure Kerr case. 

\subsection{Bumblebee model}

\begin{figure}[htbp]
\centering
\includegraphics[width=\textwidth]{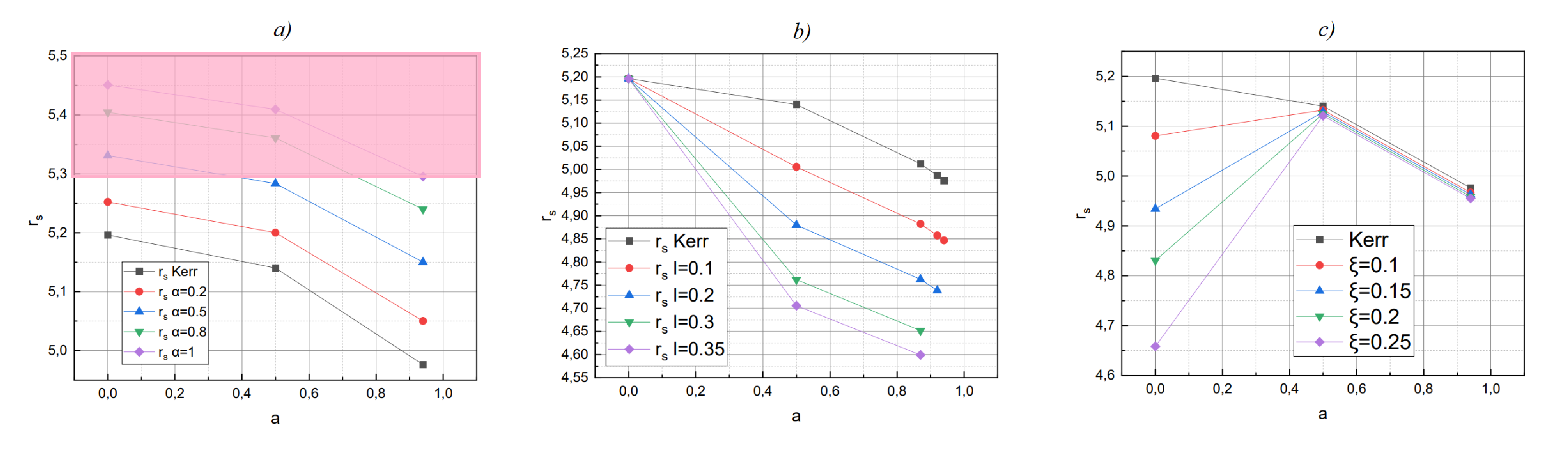}
\caption{The dependence of the shadow size $r_s$ against rotation parameter $a$ for different values of (a) $\alpha$ in Horndesky model, (b) $l$ in bumblebee model and (c) $\xi$ in scalar Gauss-Bonnet gravity. The minimal and maximal values of additional parameters are established. The inclination angle of the rotational plane is equal to $\theta_0=\frac{\pi}{6}$ (Sgr A*). The coloured region is forbidden by Sgr A* results.}
\label{rs}
\end{figure}

The next considered Kerr-like metric comes from the bumblebee model. As it was shown earlier \cite{Prokopov:2021lat} in usual bumblebee model (the same as anywhere else where the metric function $G(r)$ is equal to its Schwarzschild value) in the absence of rotation the shadow size is equal to its Schwarzs\-child value ($r_s=3\sqrt{3}M$). That is why on Fig.~\ref{sh}(b) we demonstrate only $a\neq 0$. As one can see from Fig.~\ref{rs}(b) if $l \neq 0$ the shadow size is less than in pure Kerr one. The limitations from Sgr A* allow all the values for $l$. It is importantly to note that for each value of $l$ a critical value of $a$ exists in the same manner as it was demonstrated earlier \cite{Capozziello:2023tbo}. Hence for $a=0.94$ the only value of $l$ equal to $l=0.1$ appears to be correct (for example when $l=0.2$ therefore $a_{crit}=0.92$ except $a=0.94$).

The next item to be pointed out is that the shift $D$ appears to be less than in pure Kerr case  (Fig.~\ref{D_delta}(b up)). This behaviour matches Horndesky case. The maximum occurs when $a=0.5$. Further consider distortion parameter $\delta$ (Fig.~\ref{D_delta}(b down)): when $a=0.5$ the distortion differs from the pure Kerr case and is equal approximately to $0.8-1.4\%$ depending upon $l$. When $a$ takes big values the distortion appears to be greater than in pure Kerr case (approximately $5.5\%$) and continues to increase while $l$ becomes larger and larger (up to $9.2\%$ when $l=0.2$ and $l=0.35$). Finally the distortion values for different $l$ are the same because for different $l$ has different $a_crit$. Hence if Sgr A* appear to be fast rotating BH the last result gives an additional potential to confirm the bumblebee model.

\subsection{Gauss-Bonnet scalar gravity}

The last considered solution is the rotating BH one (\ref{spin_metri_gb}) for Gauss-Bonnet scalar gravity. As it was noted earlier \cite{Zakharov:2014lqa,Prokopov:2020ewm}, when coupling parameter $\xi>0.3$ the photon sphere disappears therefore the shadow also vanishes. So we consider $\xi$ located down this limit. Fig.~\ref{sh}(c) shows the shadow profile. Consider the effective shadow size (Fig.~\ref{rs}(c)): at static case $a=0$ the shadow size decreases when $\xi$ increases. At some non-zero value of $\xi$ the effective shadow size grows faster. Further the shadow size becomes less when $\xi$ increases. Analogously to non-rotating case all allowed combinations are located inside the constrains.

\begin{figure}[htbp]
\centering
\includegraphics[width=\textwidth]{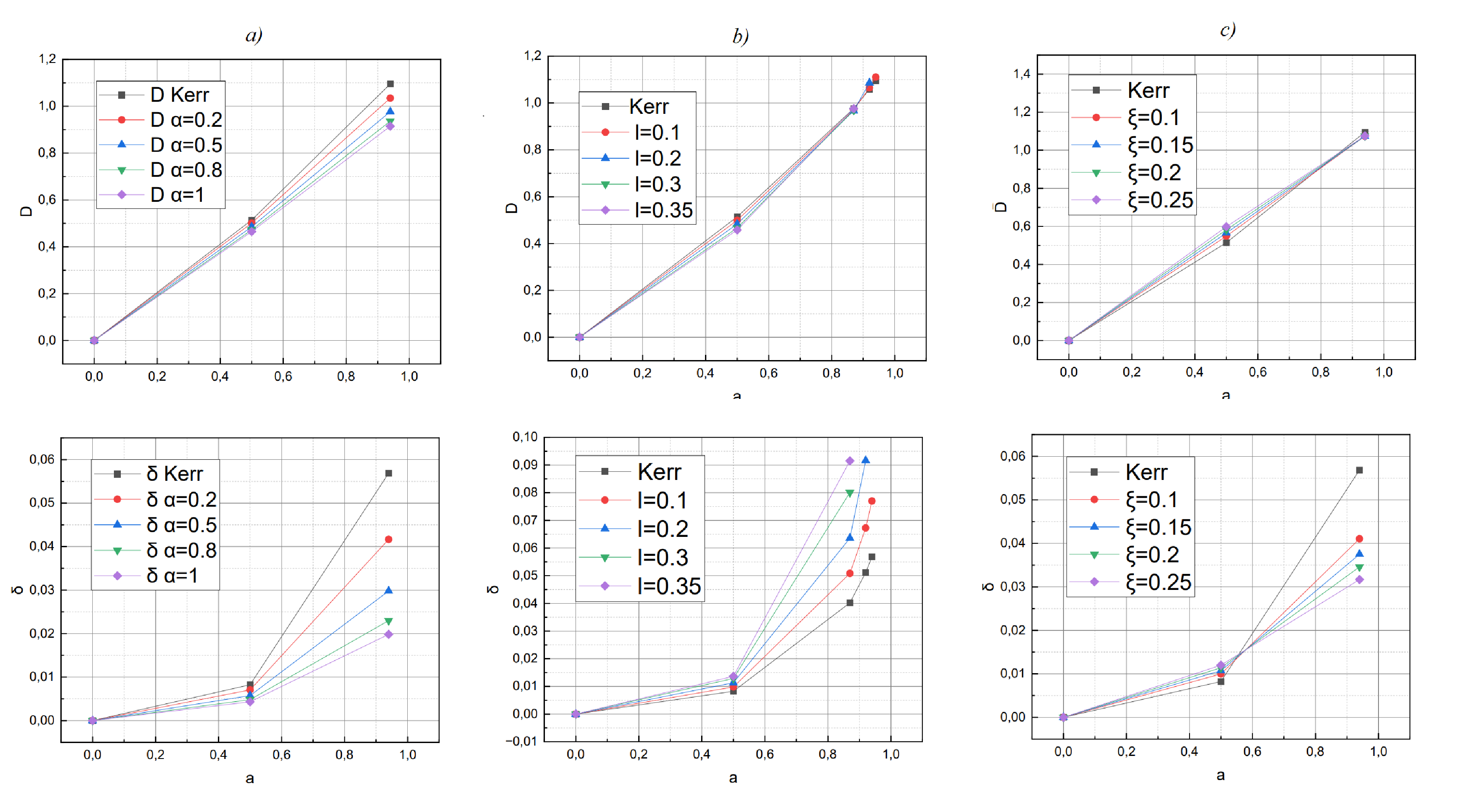}
\caption{The dependence of shift $D$ (up case) and distortion $\delta$ (down case) against rotating parameter $a$ for (a) $\alpha$ in Horndesky model, (b) $l$ in bumblebee model and (c) $\xi$ in scalar Gauss-Bonnet gravity. The minimal and maximal values of additional parameters are established. The inclination angle of the rotational plane is equal to $\theta_0=\frac{\pi}{6}$ (Sgr A*).}
\label{D_delta}
\end{figure}

Note that analogously to bumblebee model shift $D$ remains greater than in Kerr case when $a=0.5$ (Fig.~\ref{D_delta}(c up)). Also the distinction between $D$ values with zero rotation and Kerr case with $a=0.5$ is visibly greater than in previous cases. Finally consider the distortion $\delta$ behaviour (Fig.~\ref{D_delta}(c down)): when $a=0.5$ the distortion is equal about $0.8-1.2\%$ and greater than in pure Kerr case. The particular case occurs at $a=0.94$: the distortion becomes less. So the increasing of coupling $\xi$ cause decreasing of the distortion. For example for $\xi=0.25$ the distortion is approximately equal to $3.2\%$.

\section{Discussion and conclusions}

Using the improved Newman-Janis algorithm we generated rotating solutions for Horndesky model (specific partial case), bumblebee model and Gauss-Bonnet scalar gravity. As it was demonstrated  \cite{EventHorizonTelescope:2022apq,EventHorizonTelescope:2023fox} both Sgr A* and M87*  are rotating. For Sgr A* the most probable values are $a=0.5$ and $a=0.94$, for M87* it is equal to $a=0.9375$. Hence the usage of axially-symmetric BH space-time as the basis for the theoretical calculation of shadow profiles seems to be justified.

Based on obtained spinning BH metrics we calculated the shadow profiles applying the most probable configurations for Sgr A*. Therefore the inclination angle of rotation plane relatively to the observer's sight line is equal to $\theta=\pi/6$. Using EHT constrains on effective shadow size \cite{EventHorizonTelescope:2022apq} the values like $\alpha > 0.5$ at $a=0.5$ for Horndesky model were excluded. In contrast for the fast rotation at $a=0.94$ all the configurations appear to be allowed. The shadow distortion differs from Kerr one very little for $a=0.5$. For $a=0.94$ it becomes less during $\alpha$ growing (from $5.5\%$ at Kerr case till $2\%$ at $\alpha=1$). So the additional parameter acts opposite to rotation making its influence less.

The same analysis was done for bumblebee model. All the configurations appeared to be acceptable. It is important that for each $l$ its own critical value of $a$ exists. Therefore the fast rotation with $a=0.94$ is excluded for most range of $l$ values. In contrast to the previous case the distortion grows when $l$ becomes larger (up to $9.2\%$ for $l=0.2$ and $l=0.35$). So in bumblebee model the additional parameter enhances the rotational effect.

In Gauss-Bonnet scalar gravity the shadow size has minor differences from Kerr case in contrast to the static case where the difference greater even for small values of coupling parameter ($\xi=0.3$ is maximally possible value \cite{Zakharov:2014lqa}). Analogously to bumblebee model all the configurations appear to be allowed. For $a=0.5$ the distortion is greater than in pure Kerr case but for $a=0.94$ the distortion appears to be less (from $3.2\%$ at $\xi=0.25$). So the theory weakens the effect of rotation.

Generally one arrives to a conclusion that for three considered models two of them (Horndesky model and Gauss-Bonnet scalar gravity) weaken the effect of rotation and bumblebee model enhances it. This conclusion matches the previous one at non-local gravity models study: extended gravity theories by themselves correct the effect of rotation in both directions. This fact seems to be important as the accuracy of shadow images permanently increases. In addition the shadow form constraints the possible theory parameters stronger.

\acknowledgments
The work was supported by the Russian Science Foundation via grant No. 23-22-00073.


\begin{thebibliography}{99}

\bibitem{EventHorizonTelescope:2023fox}
Akiyama, K.; others. \emph{First M87 Event Horizon Telescope Results. IX. Detection of Near-horizon Circular Polarization.} \emph{Astrophys. J. Lett.} {\bf 957}, L20, (2023) \href{https://doi.org/10.3847/2041-8213/acff70} {{\ttfamily https://doi.org/10.3847/2041-8213/acff70}}, [\href{https://arxiv.org/abs/2311.10976}  {{\ttfamily arXiv:astro-ph.HE/2311.10976}}].

\bibitem{Zakharov:2005bc}
Zakharov, A.F. \emph{Black holes: Observational properties.} \emph{Int. J. Mod. Phys. A} {\bf 20}, 2321–2325, (2005) \href{https://doi.org/10.1142/S0217751X05024560} {{\ttfamily https://doi.org/10.1142/S0217751X05024560}}. 

\bibitem{Dadhich:2000am}
Dadhich, N.; Maartens, R.; Papadopoulos, P.; Rezania, V. \emph{Black holes on the brane}. \emph{Phys. Lett. B} {\bf 487}, 1–6, (2000) \href{https://doi.org/10.1016/S0370-2693(00)00798-X} {{\ttfamily https://doi.org/10.1016/S0370-2693(00)00798-X}}, [\href{https://arxiv.org/abs/hep-th/0003061}  {{\ttfamily arXiv:hep-th/0003061}}].

\bibitem{Zakharov:2014lqa}
Zakharov, A.F. \emph{Constraints on a charge in the Reissner-Nordström metric for the black hole at the Galactic Center}. \emph{Phys. Rev. D} {\bf 90}, 062007, (2014). \href{https://doi.org/10.1103/PhysRevD.90.062007} {{\ttfamily https://doi.org/10.1103/PhysRevD.90.062007}}, [\href{https://arxiv.org/abs/1407.7457}  {{\ttfamily arXiv:gr-qc/1407.7457}}].

\bibitem{Alexeyev:2018ofs}
Alexeyev, S.O.; Latosh, B.N.; Prokopov, V.A.; Emtsova, E.D. \emph{Phenomenological Extension for Tidal Charge Black Hole}. \emph{J. Exp. Theor. Phys.} {\bf 128}, 720-726 (2018)  \href{https://doi.org/10.1134/S0044451019050080} {{\ttfamily https://doi.org/10.1134/S0044451019050080}}, [\href{https://arxiv.org/abs/1812.02677}  {{\ttfamily arXiv:gr-qc/1812.02677}}].

\bibitem{Alexeyev:2022mqb}
Alexeyev, S.; Prokopov, V. \emph{Extended Gravity Constraints at Different Scales}. \emph{Universe} {\bf 8}, 283, (2022). \href{https://doi.org/10.3390/universe8050283} {{\ttfamily https://doi.org/10.3390/universe8050283}}, [\href{https://arxiv.org/abs/2205.07655}  {{\ttfamily arXiv:gr-qc/2205.07655}}].

\bibitem{Prokopov:2021lat}
Prokopov, V.A.; Alexeyev, S.O.; Zenin, O.I. \emph{Black Hole Shadows Constrain Extended Gravity}. \emph{J. Exp. Theor. Phys.} {\bf 135}, 91–99, (2022) \href{https://doi.org/10.1134/S1063776122070093} {{\ttfamily https://doi.org/10.1134/S1063776122070093}}, [\href{https://arxiv.org/abs/2107.01115}  {{\ttfamily arXiv:gr-qc/2107.01115}}];
{\it ibid} \emph{J. Exp. Theor. Phys.} {\bf 135}, 842–843, (2022). \href{https://doi.org/10.1134/S1063776122120172} {{\ttfamily https://doi.org/10.1134/S1063776122120172}}.

\bibitem{EventHorizonTelescope:2022apq}
Akiyama, K.; others. \emph{First Sagittarius A* Event Horizon Telescope Results. II. EHT and Multiwavelength Observations, Data Processing, and Calibration}. \emph{Astrophys. J. Lett.} {\bf 930}, L13, (2022). \href{https://doi.org/10.3847/2041-8213/ac6675} {{\ttfamily https://doi.org/10.3847/2041-8213/ac6675}}, [\href{https://arxiv.org/abs/2311.08679}  {{\ttfamily arXiv:astro-ph.HE/2311.08679}}].

\bibitem{EventHorizonTelescope:2019pgp}
Akiyama, K.; others. \emph{First M87 Event Horizon Telescope Results. V. Physical Origin of the Asymmetric Ring}. \emph{Astrophys. J. Lett.} {\bf 875}, L5, (2019). \href{https://doi.org/10.3847/2041-8213/ab0f43} {{\ttfamily https://doi.org/10.3847/2041-8213/ab0f43}}, [\href{https://arxiv.org/abs/1906.11242}  {{\ttfamily arXiv:astro-ph.GA/1906.11242}}].

\bibitem{Cui:2023uyb}
Cui, Y.; others. \emph{Precessing jet nozzle connecting to a spinning black hole in M87}. \emph{Nature} {\bf 621}, 711–715, (2023). \href{https://doi.org/110.1038/s41586-023-06479-6} {{\ttfamily https://doi.org/10.1038/s41586-023-06479-6}}, [\href{https://arxiv.org/abs/2310.09015}  {{\ttfamily arXiv:astro-ph.HE/2310.09015}}].

\bibitem{Prokopov:2020ewm}
Prokopov, V.; Alexeyev, S. \emph{Shadow from a rotating black hole in an extended gravity.} \emph{Int. J. Mod. Phys. A} {\bf 35}, 2040060, (2020). \href{https://doi.org/10.1142/S0217751X20400606} {{\ttfamily https://doi.org/10.1142/S0217751X20400606}}.

\bibitem{Chandrasekhar:1985kt}
Chandrasekhar, S. \emph{The mathematical theory of black holes}; Oxford University Press, NY (1985).

\bibitem{Newman:1965tw}
Newman, E.T.; Janis, A.I. \emph{Note on the Kerr spinning particle metric}. \emph{J. Math. Phys.} {\bf 6}, 915–917, (1965). \href{https://doi.org/10.1063/1.1704350} {{\ttfamily https://doi.org/10.1063/1.1704350}}.

\bibitem{Azreg-Ainou:2014aqa}
Azreg-Aïnou, M. \emph{From static to rotating to conformal static solutions: Rotating imperfect fluid wormholes with(out) electric or magnetic field.} \emph{Eur. Phys. J. C} {\bf 74}, 2865, (2014). \href{https://doi.org/10.1140/epjc/s10052-014-2865-8} {{\ttfamily https://doi.org/10.1140/epjc/s10052-014-2865-8}}, [\href{https://arxiv.org/abs/1401.4292}  {{\ttfamily arXiv:gr-qc/1401.4292}}].

\bibitem{Karas:2023uac}
Karas, V.; Stuchlik, Z. Magnetized \emph{Black Holes: Interplay between Charge and Rotation}. \emph{Universe} {\bf 9}, 267, (2023). \href{https://doi.org/10.3390/universe9060267} {{\ttfamily https://doi.org/10.3390/universe9060267}}, [\href{https://arxiv.org/abs/2306.07804}  {{\ttfamily arXiv:gr-qc/2306.07804}}].  

\bibitem{Capozziello:2023tbo}
Capozziello, S.; Zare, S.; Hassanabadi, H. \emph{Testing bumblebee gravity with global monopoles in a dark matter spike by EHT observations from M87 and Sgr A,} (2023). [\href{https://arxiv.org/abs/2311.12896}  {{\ttfamily arXiv:gr-qc/2311.12896}}] 

\bibitem{Babichev:2017guv}
Babichev, E.; Charmousis, C.; Lehébel, A. \emph{Asymptotically flat black holes in Horndeski theory and beyond}. \emph{JCAP} {\bf 04}, 027, (2017). \href{https://doi.org/10.1088/1475-7516/2017/04/027} {{\ttfamily https://doi.org/10.1088/1475-7516/2017/04/027}}, [\href{https://arxiv.org/abs/1702.01938}  {{\ttfamily arXiv:gr-qc/1702.01938}}]. 

\bibitem{Casana:2017jkc}
Casana, R.; Cavalcante, A.; Poulis, F.P.; Santos, E.B. \emph{Exact Schwarzschild-like solution in a bumblebee gravity model}. \emph{Phys. Rev. D} {\bf 97}, 104001, (2018). \href{https://doi.org/10.1103/PhysRevD.97.104001} {{\ttfamily https://doi.org/10.1103/PhysRevD.97.104001}}, [\href{https://arxiv.org/abs/1711.02273}  {{\ttfamily arXiv:gr-qc/1711.02273}}].

\bibitem{Yunes:2011we}
Yunes, N.; Stein, L.C. \emph{Non-Spinning Black Holes in Alternative Theories of Gravity}. \emph{Phys. Rev. D} {\bf 83}, 104002, (2011). \href{https://doi.org/10.1103/PhysRevD.83.104002} {{\ttfamily https://doi.org/10.1103/PhysRevD.83.104002}}, [\href{https://arxiv.org/abs/1101.2921}  {{\ttfamily arXiv:gr-qc/1101.2921}}]. 

\bibitem{c02}
Alexeyev, S., B.A.; Zenin, O. \emph{Non-local gravitational corrections in black hole shadow images}. \emph{Zh. Exp. Theor. Phys.} {\bf 165}, 508–515, (2024). \href{https://doi.org/10.31857/S0044451024040059} {{\ttfamily https://doi.org/10.31857/S0044451024040059}}, {\it (in Russian)} [\href{https://arxiv.org/abs/2404.16079}  {{\ttfamily arXiv:gr-qc/2404.16079}}].

\end{thebibliography}
\end{document}